# Amorphous Germanium as a Promising Anode Material for Sodium Ion Batteries: A First Principle Study

Vidushi Sharma[1], Kamalika Ghatak[1], Dibakar Datta*

Department of Mechanical and Industrial Engineering

New Jersey Institute of Technology, Newark, NJ 07103, USA

[1] Equal contribution

*Corresponding author

Dibakar Datta; Email – dibakar.datta@njit.edu ; Phone – 973 596 3647

**Abstract**

The abundance of Sodium (Na), its low-cost, and low reduction potential provide a lucrative inexpensive, safe, and environmentally benign alternative to Lithium Ion Batteries (LIBs). The significant challenges in advancing Sodium Ion Battery (NIB) technologies lies in finding the better electrode materials. Experimental investigations revealed the real potency of Germanium (Ge) as suitable anode materials for NIBs. However, a systematic atomistic study is necessary to understand the fundamental aspects of capacity-voltage correlation, microstructural changes of Ge, as well as diffusion kinetics. We, therefore, performed the Density Functional Theory (DFT) and Ab Initio Molecular Dynamics (AIMD) simulation to investigate the sodiation-desodiation kinetics in Germanium-Sodium system ($Na_{64}Ge_{64}$). We analyzed the intercalation potential and capacity correlation for intermediate equilibrium structures and compared our data with the experimental results. Effect of sodiation on inter-atomic distances within Na-Ge system is analyzed by means of Pair Correlation Function (PCF). This provides insight into possible microstructural changes taking place during sodiation of amorphous Ge (*a*-Ge). We further investigated the diffusivity of sodium in *a*-Ge electrode material and analyzed the volume expansion trend for $Na_{64}Ge_{64}$ electrode system. Our computational results provide the fundamental insight into the atomic scale and help experimentalists design Ge based NIBs for real-life applications.

**Keywords:** Density Functional Theory; Ab initio Molecular Dynamics; Sodium Ion Batteries; Energy; Electrodes; Germanium

1. INTRODUCTION

The periodic increment in the global energy demand is keeping researchers busy recognizing the potential energy materials for safe, efficient and inexpensive energy storage. With the imminent exhaustion of fossil fuel and its subsequent environmental consequences, there has been enormous demand for eco-friendly, renewable, cheap, and portable secondary battery that could provide energy storage for variable applications. In past few decades, Li-ion batteries (LIBs) have ruled the market as leading battery technology owing to their high-energy density, long cycle life, and lightweight.[1-3] They have been the primary source of power in portable electronic devices and electric vehicles. However, with the ever-increasing demand for LIBs, there have been concerns about depleting Lithium (Li) reserves and the cost associated with this technology. Alternatively, Sodium (Na) being the fourth most abundant element on earth, offers a potential rechargeable electrochemical energy storage (EES) in the form of Na-ion battery (NIB).[4-7] Moreover, Na is cheaper and has its redox potential closer to that of Li ($E^o_{(Na+/Na)}$ = -2.71V).[8] All these advantages make NIB a real alternative to LIB. Although Na is heavier than Li, resulting comparatively lower energy density than LIBs, it is a small penalty to pay for safer, cheaper, abundant, and rechargeable energy storage system.

Concerning battery components and electrical storage mechanisms, NIB is very similar to LIB; the only exception lies in their ion carriers. The difference in ion carriers, i.e., Na being bigger and heavier as compared to Li, is what makes a difference in intercalation mechanism and transport phenomenon in the battery systems.[9] Na, being in the same group as Li and possessing similar chemical behavior, shows identical diffusion kinetics and comparable diffusion barriers as that of its Li counterpart while combined with solid state host materials.[10] Similar charge/discharge profiles are reported for NIBs and LIBs except for some different plateau observed in NIB due to Na size induced strain effects.[4] Na being bigger in size offers sluggish transport, and therefore, it has been reported that LIB electrodes tend to exhibit better cycle stability. Since capacity and cell cycle are primarily determined by electrodes, it becomes essential to search proper electrode materials for NIB to become as efficient as LIB.

Much advancement has already been made in studying cathode materials for NIB.[11-13] Analogous to LIB, sodium layered oxide compounds, and polyanion compounds have been extensively studied in recent years.[14] Most of these studies are based on improved designing of the host species to accommodate large Na ions so that minimal structural change takes place during the charge/discharge cycle. To the best of our knowledge, on the contrary to the positive electrode research, negative electrodes for NIB are less explored and are still open to proper investigation. The existing studies lack the detailed understanding of the bulk and interfacial properties of the systems. An efficient anode material for NIB with significant capacity, charge/discharge rate, and enhanced cycle life is still a challenge. Initially, crystalline carbonaceous anode such as graphite was investigated as a potential electrode material for NIB owing to their success as the anode in LIB,[15] but the degree of Na intercalation through graphite framework was found to be negligible due to its larger diameter.[16-18] To overcome this, Wen et al.[19] put forward expanded graphite with large interlayer spacing as an anode material for NIB that could ease $Na^+$ ion intercalation and provides the decent capacity of 150 mAhg$^{-1}$. However, such structures are synthesized via oxidation of Graphene followed by partial reduction which results in the residual oxygen-containing groups in the interlayers. These oxygen-containing groups hinder $Na^+$ ion intercalation and therefore affecting the overall efficiency of the electrode.[19] Henceforth, focus shifted towards the disordered carbonaceous materials such as hard carbon and metal oxides such as titanium-based oxides[20-21]. These materials were found to exhibit better reversibility during electrochemical processes in NIBs and provide comparatively high specific capacities.

Table 1 lists some of the anode materials with their specific capacities as reported in the respective literature. It has been studied that heteroatom doping in carbonaceous materials significantly improves its specific capacity by creating defect sites which eventually facilitates $Na^+$ absorption and charge transfer process.[22] For example, sulfur(S)-doped carbon exhibited very high specific capacity of 516 mAhg$^{-1}$. Transition Metal Oxides (TMOs) and Transition Metal Sulfides (TMSs) have also been reported as potential anodes for NIBs due to their high capacities.[23] However, their sodiation/desodiation mechanism is based on conversion reaction which leads to significant structural changes in the electrode and substantial volume expansions. Further, group 14 and 15 elements tend to form alloys with Na, which exhibit very high capacities (ranging from 350 – 850 mAhg$^{-1}$ in case of group 14 elements and 385 – 2560 mAhg$^{-1}$ in case of group 15 elements).[24] Despite high capacities, their practical application is currently limited as they have lower cycle life and undergo high-volume expansion.[17, 25] These alloys with high storage capacity can be exploited further to improve the charge-discharge cycle life and to control volume expansion.

*Table 1*: Anode materials for NIB with their investigated capacities.

| S. no. | Anode material | Capacity in mAhg$^{-1}$ |
|---|---|---|
| 1. | Hard Carbon | 300[26] |
| 2. | Expanded Graphite | 150[19] |
| 3. | Reduced Graphene oxide | 141[27] |
| 4. | N doped porous nanofibres | 212[28] |
| 5. | S doped Carbon | 516[22] |
| 6. | TiNbO$_2$ | 160[29] |
| 7. | S-TiO$_2$ | 320[30] |
| 8. | Na$_2$Ti$_3$O$_7$ | 89[31] |
| 9. | Na$_2$Ti$_3$O$_{15}$ | 258[32] |
| 10. | *a*-Si | 725[33] |
| 11. | *a*-Ge | 369[34] |
| 12. | Sn | 847[35] |

High theoretical capacities of group 15 elements and higher group 14 elements such as Tin (Sn) for NIB has encouraged researchers to investigate them in more detail in terms of intercalation voltage profile[36], diffusivity[33], elasticity[25] etc. The resultant high specific capacities are due to their ability to take up more than one Na per metal atom. Volumetric energy densities of these elements can go up to 3.5 Whcc$^{-1}$.[24] This property may attribute to their very high capacities, but is also a major cause for very high volume expansion (greater than 150%) and microstructural changes in these anode materials. Repeated sodiation in such electrodes causes self-pulverization which ultimately leads to fracture with the loss of cycle life.[37] In such a scenario, lower group 14 elements such as Silicon (Si) and Germanium (Ge) seem to be much promising options.

Si has been one of the most extensively studied anode material for NIB due to its enormous success in case of LIBs. All forms of Si have been probed for their potential as the anode in NIBs. Initially, crystalline Si (*c*-Si) was investigated for Na diffusion kinetics, and it was reported that in spite of being chemically similar to Li, Na intercalation in *c*-Si is limited because of its bigger size. Bulk Si has been computationally investigated to have a very high energy barrier (1.41 eV) for Na diffusion.[38] Kulish et al.[39] investigated layered Si such as polysilane as an anode for NIB. They showed by means of DFT and Nudge Elastic Band (NEB) method that energy barrier for Na diffusion in such compounds gets reduced to 0.41 eV, but these materials exhibit low capacity of 279 mAhg$^{-1}$.[39] Failure of layered Si motivated the intense investigations of *a*-Si, where again Na displayed very low diffusivity that makes it an impractical choice.[33]

All of this trial experimentation leaves researchers with the most viable option of considering Ge, as the potential host material for Na$^+$ ions due to its vast similarities with the Li-Si system. Like Si, Ge too takes up one Na per atom to form Na-Ge alloy.[24] Theoretical capacity for the Na-Ge system have been reported to be 369 mAhg$^{-1}$, which may be lower than other alloys but is still better than C-based and TMOs/TMSs anodes.[34] Furthermore, Na-Ge anode is known to undergo comparatively lower volume expansion than other available alternatives. This makes Ge a promising anode material for NIB. However, unlike Si, Ge is not well studied and requires thorough investigations from both the experimental and theoretical point of view in order to gain more insights. Recently, an experimental investigation by Baggetto *et* al., reported *a*-Ge thin film as a promising anode material for NIBs.[40] To the best of our knowledge, not much theoretical studies are done on Na-Ge system leaving the scope of the further insightful atomistic investigation to fully comprehend its equilibrium intercalation voltage curve, diffusion kinetics, and possible structural changes. In this study, the first principle Density Functional Theory (DFT) is employed to analyze intercalation potential vs. capacity curve and the simulation results are compared with the experimental data. The Ab initio Molecular Dynamics (AIMD) method is incorporated to investigate the diffusion kinetics of single Na atom in *a*-Ge. AIMD simulation draws a more realistic picture of intermixing at atomistic level and helps to trace Na trajectory through $a$-Ge. Our study manifests higher diffusivity of Na in $a$-Ge as compared to other known anode materials. Moreover, a thorough analysis of Pair Correlation Function (PCF) reveals the insight into the microstructural changes at different stages of sodiation into germanium.

## 2. COMPUTATIONAL DETAILS

The first principle calculations were done using Vienna Ab Initio Simulation Package (VASP).[41] The optimized equilibrium structures were obtained via DFT as implemented in VASP. PAW pseudopotentials[42] were taken for the inert core electrons, and valence electrons were represented by plane-wave basis set. The GGA, with the PBE (Perdew-Burke-Ernzerhof) exchange-correlation functionals were taken into account.[43] All DFT relaxation includes force, geometric, volume, and cell shape relaxations until the minimum energy criteria of $1.0 \times 10^{-4}$ eV was met. All the internal coordinates are relaxed until the Hellmann-Feynman forces are less than 0.02 eV/Å. Plane wave cutoff for all the calculations was taken as 230 eV following high precision convergence as described in VASP and was tested accordingly. The Brillouin zone was sampled with 1×2×1 mesh in Monkhorst pack grid.

Initially, we investigated the intercalation of Na through $c$-$Ge_{64}$ at various elevated temperatures (below the melting point of Ge). We intended to model alloy anode starting with $c$-$Ge_{64}$. However, we discovered that Na, due to its larger diameter, does not intercalate into the $c$-$Ge_{64}$ lattice. So, $a$-Ge was considered for our study. In order to model $a$-$Ge_{64}$, we started from $c$-$Ge_{64}$ and generated $a$-$Ge_{64}$ via computational quenching process. The quenching process is a combined AIMD and DFT relaxations involving heating, cooling and equilibration for significant amount of time steps (in this case 5000MD time steps with 1 fs time interval). In our study, the highest and lowest temperatures for the AIMD run were 5000K and 298 K (room temperature). Finally, $a$-$Ge_{64}$ was obtained via further DFT optimization of the room temperature AIMD simulated lowest energy (local minima) $Ge_{64}$ structure. Amorphous phase of the generated structure was confirmed by taking Radial Distribution Function (RDF) at every step of the process.

In this study, we have considered Na-Ge alloy in 1:1 ratio. To be more specific, we have considered the $Na_{64}Ge_{64}$ system in order to calculate intercalation potential vs. capacity followed by the consecutive desodiation leading towards the $Ge_{64}$ system. The reason behind the consideration of this particular stoichiometry is the similar volumetric energy density of Na-Ge system with Li-Si system. Later has been studied with the 1:1 atomic ratio which motivated us to do the same.[44] Previous DFT study to model $a$-Ge as an anode for NIB employed Na:Ge ratio greater than 1 which is not experimentally found.[45] Their reported theoretical specific capacity value is 576 mAhg$^{-1}$, which is much greater than the experimentally feasible capacity for Ge anode.[46] The ratio taken in the present study gave specific capacity (369 mAhg$^{-1}$) for $a$-Ge anode closer to experimentally and theoretically investigated capacities.[34, 40] The formation of mixed $Na_{64}Ge_{64}$ alloy failed initially when $c$-$Ge_{64}$ was considered as the precursor. Alternatively, consideration of $a$-$Ge_{64}$ as the precursor in this mixing process facilitates the Na penetration through the amorphous network. $Na_{64}Ge_{64}$ mixed system was generated via incorporation of 64 Na atoms into $a$-$Ge_{64}$ using AIMD simulation. Again RDF was used to determine the phase of the final alloy mixture. Figure 1 depicts an overall picture of the entire process.

To compute the intercalation voltage profile (V vs. Na/Na+) for Na-Ge binary anode system, several intermediate structures with Na concentrations varying by 6.25% were taken into account and optimized by DFT. In a system $Na_nGe_{64}$, $n$ is the number of Na atoms inserted in the computational cell and varies from 0 to 64. Specific capacity for Na-Ge system was calculated using the formula

$$C = \frac{nF}{3.6 \times M} \quad (1)$$

where **C** is the specific capacity (mAhg$^{-1}$), ***n*** is the number of charge carriers (Na in this case), **F** is Faraday's constant, and **M** is molecular weight of the active material used. In our present study, we have considered $a$-$Ge_{64}$ as the active anode material. The energy minimization calculations provide information about the formation energy of the system ($\Delta E_f$) and subsequently sodiation potential **V(*n*)** which is given by

$$V = \frac{\Delta G_f}{z_e F} \quad (2)$$

where, $z_e$ is the charge carried by Na in the electrolyte, and $\Delta G_f$ is the change in Gibbs free energy given by

$$\Delta G_f = \Delta E_f + P\Delta V_f - T\Delta S_f \quad (3)$$

The pressure and entropy components in the above equation can be neglected as they are of very small order as compared to $\Delta E_f$. This makes free energy equivalent to formation energy defined by

$$\Delta E_f = E_{Na_nGe_{64}} - (nE_{Na} + E_{Ge}) \quad (4)$$

where $\Delta E_f$ are the formation energies of the systems with varying Na concentrations, $E_{Na_nGe_{64}}$ is the energy of systems with varying Na concentrations, $E_{Ge}$ and $E_{Na}$ are respective energies of anode system without Na and energy of single Na atom which in our case was calculated to be -1.307 eV.

In order to account for the intermediate phase changes in Na-Ge system, Pair Correlation Function (PCF), *g(r)* were calculated for different AIMD trajectories of different Na concentrations in *a*-$Ge_{64}$. These AIMD run were carried out for different $Na_nGe_{64}$ species for 5000 MD steps with the time interval of 3 femtosecond (fs). PCF throws light on inter-atomic distances between the atoms throughout the process. Such an analysis helps in determining amorphous or crystalline nature of the system. PCF, *g(r)* for homo-atomic pairs in our system (Ge-Ge and Na-Na) were calculated and plotted. Here, the *g(r)* is the second order correlation function $g^2(r_{12})$, where $r_{12}$ (=$r_2$-$r_1$) represents the distance between two atoms. The mathematical formula for PCF is given below, where (*Z-1*) represents the number of nearest pairs and $\rho$ is the probability density.

$$\int_{r=0}^{\infty} \rho\, g(r) \cdot 4\pi r^2\, dr = Z - 1 \quad (5)$$

Next, we investigated diffusivity of single Na in equilibrated $a$-Ge$_{64}$. Diffusion of single Na through $a$-Ge$_{64}$ was determined by stimulating intercalation of Na atom at three elevated temperature conditions (1100 K, 950 K, 800 K) using AIMD for 15000 MD steps with the timestep of 1fs. These AIMD trajectories were then used to compute mean squared displacement (**MSD**) for the determination of diffusion trajectory and diffusivity of Na atom in the Ge$_{64}$ system. The MSD was computed by

$$\mathbf{MSD} = \frac{1}{N}\sum_{n=1}^{N}(x(t) - x_n(0))^2 \qquad (6)$$

where $N$ is the number of particles to be averaged and $x$ represents positions of the particle at different time frames. Further, Einstein equation **MSD = 6Dt** was taken into consideration in order to compute diffusivity of Na in Ge at various temperatures, where **D** is the diffusivity to be calculated and **t** is the time taken. In general, these AIMD simulations are done at high temperatures to reduce the simulation time by accelerating the rate of reaction. Diffusivities calculated at elevated temperatures were then extrapolated to room temperature (~300K) using Arrhenius equation:

$$\mathbf{D} = \mathbf{D_0}\ e^{\frac{-E_b}{kT}} \qquad (7)$$

Where, **D$_0$** is pre-exponential factor, **E$_b$** symbolizes the energy barrier, and **k** and **T** represent the Boltzmann constant and simulation temperature respectively.[44]

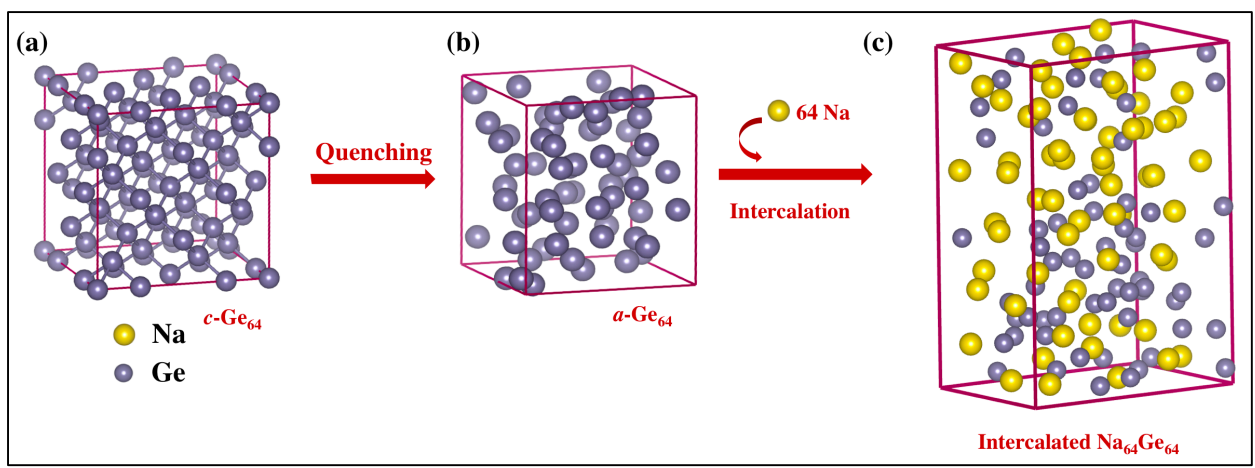

**Figure 1**: (a) Optimized structure of the initial crystalline $Ge_{64}$ ($c$-$Ge_{64}$) (b) Optimized amorphous $Ge_{64}$ ($a$-$Ge_{64}$) system obtained via quenching process of $c$-$Ge_{64}$ (c) Optimized structure of $Na_{64}Ge_{64}$ obtained by intercalating $Na_{64}$ into $a$-$Ge_{64}$ via AIMD simulation.

## 3. RESULTS AND DISCUSSION

Initial investigation of $c$-$Ge_{64}$ as alloy anode material for NIB was not successful as Na does not intercalate into the $c$-$Ge_{64}$ lattice due to its larger diameter. This is because of the fact that $c$-$Ge_{64}$ has similar crystal lattice structure to $c$-Si (Diamond FCC). Both the crystal structures have lattice constants very close in value (565.75 ppm for $c$-Ge and 543.09 ppm for $c$-Si) while atomic radii of Ge is significantly larger than Si. This makes $c$-Ge even less suitable material for Na intercalation than $c$-Si which has already been reported to have very limited Na intercalating tendency.[38] $a$-$Ge_{64}$ was then modeled starting from $c$-$Ge_{64}$ via computational quenching process as described in computational section. One of the important observation was that the Na atoms were mixed easily with $a$-$Ge_{64}$ than with $c$-$Ge_{64}$. At the end of the room temperature AIMD simulation followed by the DFT optimization, we were able to achieve completely mixed $Na_{64}Ge_{64}$ alloy.

**Intercalation Voltage and Volume Changes**

The performance of any battery largely depends upon the efficiency of the electrodes to let the charge carriers intercalate efficiently. The electrochemical potential of NIB depends upon the smooth movement of Na atom through the host material (Ge in this present study), and it changes with the concentration of Na atoms which has a direct relationship with the capacity of the electrode system (see equation 1). Using DFT, equilibrium curve between charge and discharge curve can be obtained. Tracing such equilibrium Voltage vs. Capacity curve is the essential part of an electrochemical process as it not only gives the idea about battery performance but also provides insight into significant structural or phase change phenomenon happening during the process. In order to compute intercalation potential, we started from the $Na_{64}Ge_{64}$ system and obtained the intermediate $Na_nGe_{64}$ structures by removing 4 Na atoms (~6.25%) at a time, where $n$ varies from 0 to 64. After modeling the initial structures for all of the $Na_nGe_{64}$ phases, they were fully relaxed until the energy minimized structures were obtained. The potential value of each of these corresponding structures was computed by implementing the equation 2. Figure 2(a) depicts the potential curve obtained for intermediate $Na_nGe_{64}$ phases for the Na-Ge system. In Figure 2 (a), it is seen that voltage profile ranges from 2.08 V to 0.48 V. As Na concentration increases, there is a gradual potential drop until $n = 16$ (92 mAhg$^{-1}$ capacity) which then reduces upon further Na incorporation. There is a rise in voltage vs capacity curve at around 100 mAhg$^{-1}$ (when $n = 20$). Such 'voltage spike' is seen when there is a small energy difference in total energy of similar neighboring phases. When Na concentration reaches up to 50% ($n = 32$, 150 mAhg$^{-1}$ capacity), potential drop seems to become insignificant, and a plateau-like curve is noticeable henceforth. These kinds of the plateau are also indicative of the existence of single-phase system inside the battery during the electrochemical process. This stability of single phase (for Na > 50%), i.e., the plateau-like curve is in contrast with initial voltage jumps and spikes in the curve, where Na concentration ($n = 20$) is less. Therefore, we can say that structural phase instability is observed at low Na concentration.

In general, the addition of Na atoms should lead to volume expansion due to large atomic radii of Na atoms. However, during the DFT relaxation at different Na concentrations, a little different trend was observed regarding the volume change. It is essential to note here that upon addition of a small percentage of Na, there was initially reduction in cell volume up to 19% of Na content ($n = 12$), which increased later on. This initial volume reduction and instability of equilibrium Voltage vs Capacity curve (see figure 2(a)) for n < 20, strongly imply that chemical interactions are going on upon insertion of a small number of Na atoms in the $a$-Ge$_{64}$ cell.

Addition of Na atoms cause rapid bond formation among neighboring atoms which is the probable reason behind the volume contractions. On the other hand, with the increasing concentration of Na ($n>16$), the compressive stresses in battery anode also increase. As a result, the system acquires amorphousness by breaking these bonds to relieve these stresses. There is a subsequent Brownian motion in the system causing mixing of Na atoms, and hence there is overall volume expansion noted for the system. This hypothesis is supported by Na-Ge interatomic distances measured for various Na concentrations. There were significant variations seen for short-range atomic distances for small Na concentrations. The nearest average neighboring distance for 6% Na content ($n=4$) was 3.327 Å, while for 19% Na content, it was 3.193 Å. Beyond 19%, there is a steady increment in the average Na-Ge neighboring distances upon further Na addition. Finally, in a fully sodiated state ($Na_{64}Ge_{64}$), cell volume expanded by 149.51% which is less than the other alloying compounds studied as a potential electrode material such as Sb (390%), Phosphorous (490%), Sn (440%), Pb (500%).[33, 47] On the other hand, single Na atom occupies around 31.9Å$^3$ volume in $a$-$Ge_{64}$ and increases the volume of $a$-$Ge_{64}$ by 2.23%. Figure 2(b) illustrates DFT derived equilibrium curve overlaid on the experimental charge/discharge curve for the first cycle in $a$-Ge taken from the Galvanostatic Intermittent Titration Technique (GITT) investigation by Baggetto et al.[40] This GITT investigation is the constant current and quasi-equilibrium measurements. DFT, as implemented in VASP, deals with equilibrium thermodynamics and thus DFT derived curve represents the equilibrium configurations between the experimental charge and discharge profile. As per the expectation, our theoretical curve falls in the middle potential values, i.e., close to the mean voltages from experimental charge/discharge. This curve provides a good match with the reported GITT measurements. In experimental discharge curve, there is sudden potential drop until 0.9 V, which is close to our calculated curve.[40] Eventually, the possibility of phase instability during lower Na concentration as seen in our equilibrium curve, motivated us to perform PCF analysis for each modeled Na-Ge system.

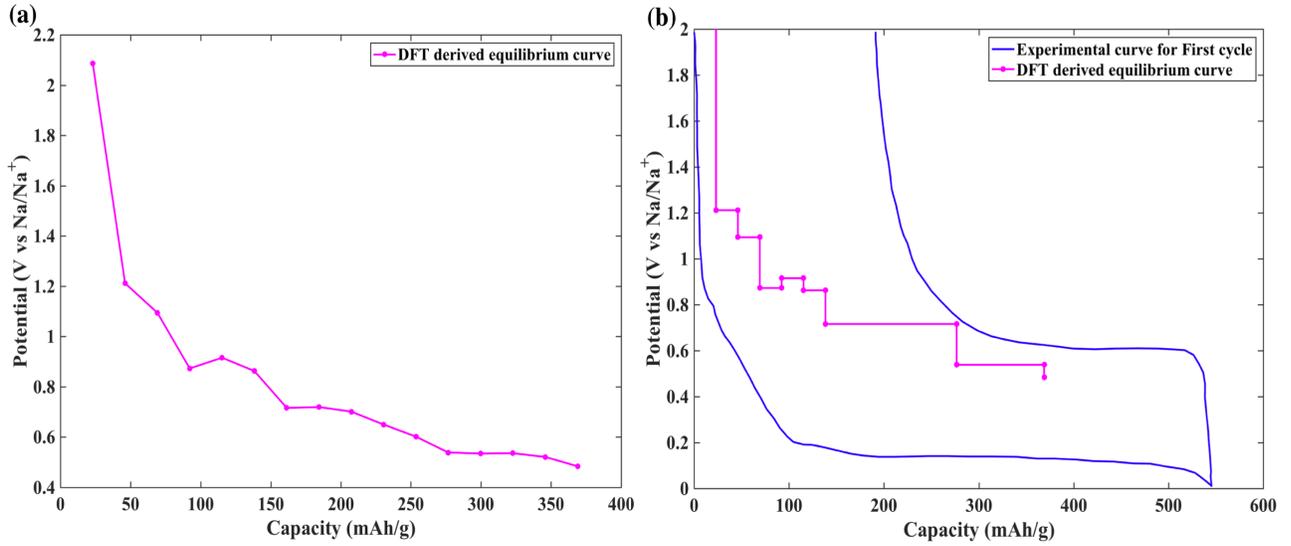

***Figure 2***: (***a***) DFT calculated Voltage vs. Capacity curve for Sodium Intercalation in *a*-Ge and (***b***) DFT derived equilibrium curve for intermediate stages of intercalated Na overlaid on experimental curve for Sodiation and Desodiation for first cycle in *a*-Ge.

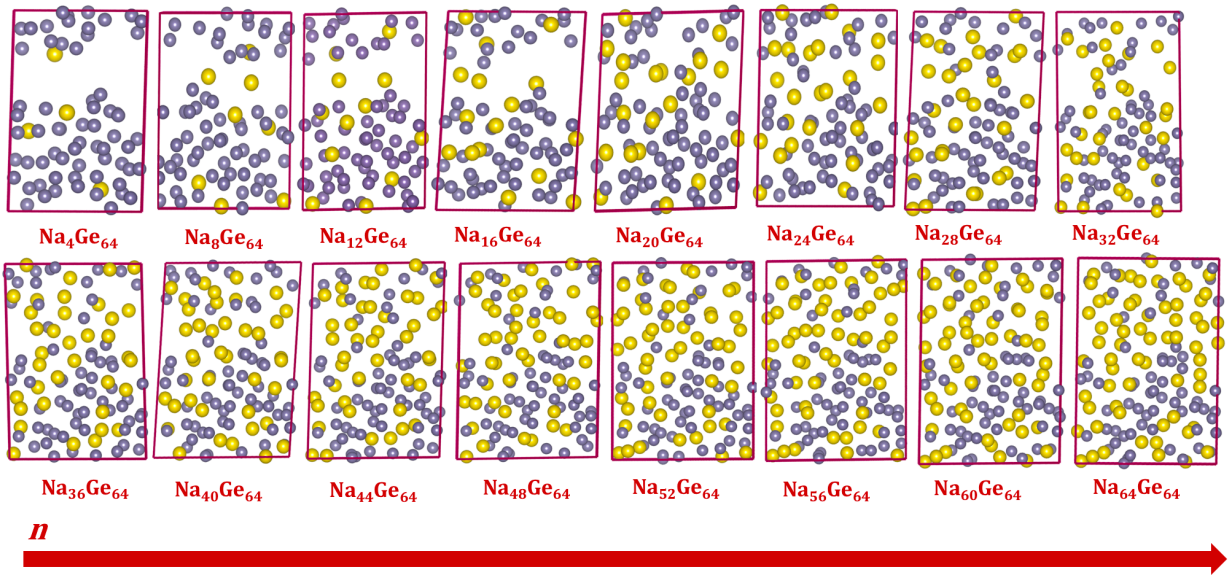

***Figure 3***: Optimized Na$_n$Ge$_{64}$ structures in between *a*-Ge$_{64}$ and Na$_{64}$Ge$_6$ where *n* refers to the number of Na atoms and varies from 4 to 64.

**PCF Analysis**

The evolution of phase and structural changes of the electrode during intermediate stages were analyzed by evaluating PCF, *g(r)* of atomic pairs (Ge-Ge and Na-Na) in the system with varying Na concentrations. AIMD simulations were performed for Na-Ge systems with intermediate Na concentrations (6%, 19%, 31%, 50%, 69% and 88%). These concentrations were randomly chosen to get an idea of phase evolution throughout the sodiation cycle. *g(r)* was calculated using previously described equation 5 and plotted against neighboring distance *(r)*. Figure 4(a) illustrates Ge-Ge PCF at 1000 MD time step for different Na concentrations. A single prominent peak at 2.8 Å was observed for all of the frames, which is close to the expected Ge-Ge first neighboring distance in the amorphous state (2.4 Å). No other significant peak was detected beyond the range of first neighboring distances. The absence of other peaks clearly signifies the amorphous nature of the Ge-Ge pairs. Highest *g(r)* value was obtained for the frame with 50% Na concentration. It is quite visible from figure 4 (a) that Ge displays amorphous nature throughout intermediate Na-Ge phases. This result is similar to previously reported analysis[46], but so far not much attention has been given to the effect of Na-Na pairs on structural changes in Ge anode upon sodiation. This inspired us to perform the PCF analysis of Na-Na pairs in order to account for their phase change activity during initial sodiation. Therefore, we calculated *g(r)* for Na-Na at 1000 MD time step for Na concentrations varying from 6% to 88 % (figure 3(b)). 6% Na showed a high peak at 5.6 Å. The nearest neighboring distance between Na atoms in *c*-Na is 3.7 Å which matched with the molecular structure with 19% Na concentration. With increasing Na concentration, nearest neighboring distance decreases indicating mixing of the system. Significant loss of peak intensity was also detected with increasing Na concentration indicating loss of crystallinity and complete amorphization. At 88% Na concentration, no prominent peak was observed, and it became almost constant suggesting total amorphous phase. These results imply that single phase detected (plateau-like curve in Figure 2(a)) for higher Na concentration (beyond 50%) is representative of the amorphous phase. On the other hand, in case of lower Na concentrations, brief crystalline phases and volume contraction were detected due to their inter-atomic interactions. In case of NIB, such phase transitions are pretty common and are typically induced by biphasic reactions. Previous literature reported the similar biphasic nature for several other alloying anode materials[24-25] and one of the most well-studied examples is the Na-Sn system.[37] However, to the best of our knowledge, no such report exists in Na-Ge system till date.

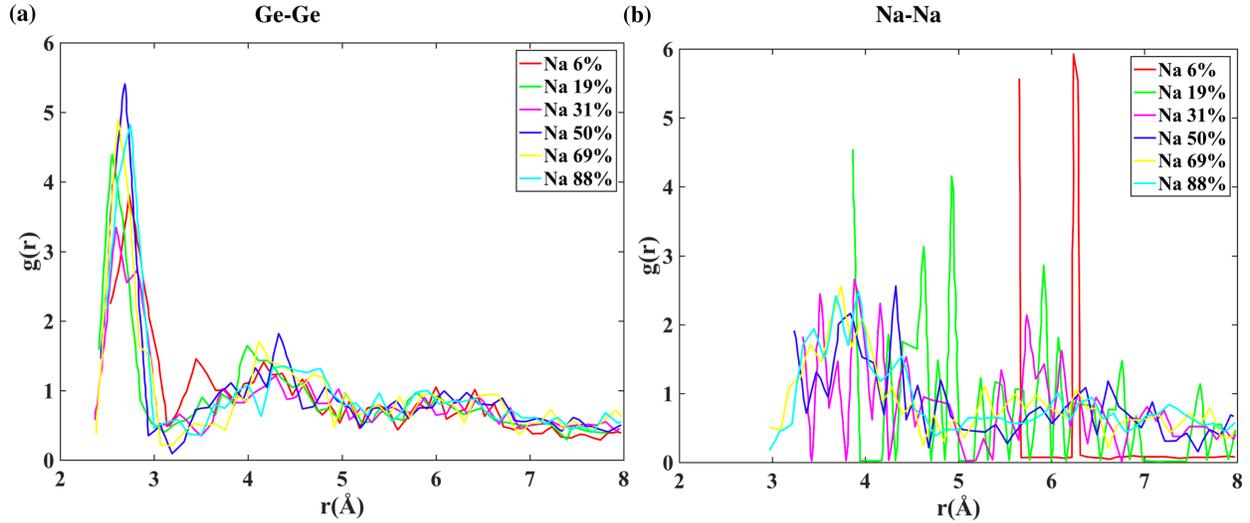

*Figure 4:* (a) Ge-Ge PCF at 1000 MD step for sodium concentrations varying from 6% to 88%. (b) Na-Na PCF at 1000 MD step for sodium concentrations varying from 6% to 88 %.

**Diffusivity of Na in amorphous Ge**

Another critical feature that is important for battery performance is fast diffusion kinetics of Na in anode material. Diffusivity determines the ability of Na to mix with the anode material and charge/discharge rates of the battery. Importance of diffusivity study is evident from the fact that Si, being the most promising anode material for NIB due to its high capacity and low volume expansion, cannot be practically applied to NIB owing to the sluggish diffusivity of Na in Si. [38] As reported, Na diffusion encounters energy barrier of 1.41 eV in bulk-Si. Therefore, it becomes crucial to extensively study the diffusion kinetics of Na in Ge, as it is one of the essential criteria concerning choosing a suitable anode. Diffusivity of Na in Ge was determined by calculating average **MSD** (equation 6) of single Na in bulk *a*-$Ge_{64}$ as a function of MD time step at different temperatures. There is a previous report[46] on Na diffusivity in *a*-Ge at very high temperatures (above the melting point of Ge). However, such high-temperature calculations are not suitable for a proper fundamental understanding of the diffusion kinetics[44] occurring in batteries. In our present study, the simulation temperatures are below the melting point of Ge (1200 K). This is in accordance with the previous diffusion studies.[44] Figure 5(a) depicts mean squared displacement of Na with respect to MD time steps for different temperatures namely

1100 K (red), 950 K (green) and 800 K (magenta). The linear increment in MSD plot with time and the increasing temperature is evident from our calculation. It was observed during AIMD run that diffusion of Na in *a*-Ge was slower at first and impeded by Na size, but with increasing time step, it shows gradual increment. The Einstein equation was implemented to calculate the diffusivity of Na in Ge (**$D_{Na}$**) at various temperatures (Table 2), which was then extrapolated to room temperature (300K) using Arrhenius equation for diffusivity (equation 7).[44] We calculated **$D_{Na}$** value for the $Na_1Ge_{64}$ system to be about $4.876 \times 10^{-9}$ $cm^2/s$ at 300 K. This magnitude is comparable to reported data for Lithium diffusivity in graphite anodes which is of about $10^{-8}$ to $10^{-10}$ $cm^2s^{-1}$.[48] This shows that Na atoms diffuse in Ge faster than in other alloying anode materials by about one order of magnitude.[33] An Arrhenius plot (lnD vs 1000/T) for $Na_1Ge_{64}$ system is shown in figure 5(b). Such plot provides an estimate of migration activation energy for Na in *a*-Ge. Migration activation energy was derived to be 0.709 eV, which compares well with currently applicable LIBs [48] and thus holds a promise of good charge/discharge rate in Na-Ge anode based NIB. This activation energy for Na in *a*-Ge is calculated from single atom diffusion model. However, diffusivity is expected to increase with Na concentration as it is established that dopant-dopant interaction (Na-Na in this case) significantly lowers the energy barrier by causing additional relaxation of surrounding atoms.[38] So, one can expect higher overall diffusivity in case of the $Na_{64}Ge_{64}$ system due to the presence of higher Na concentration during the process of charging and discharging. However, the theoretical activation energy for Na hopping in *c*-Ge has been reported to be 1.5 eV [10]**,** which is very high and thus explains the impedance Na faced when we tried Na diffusion in *c*-Ge which is apparently due to the large size of Na**.**

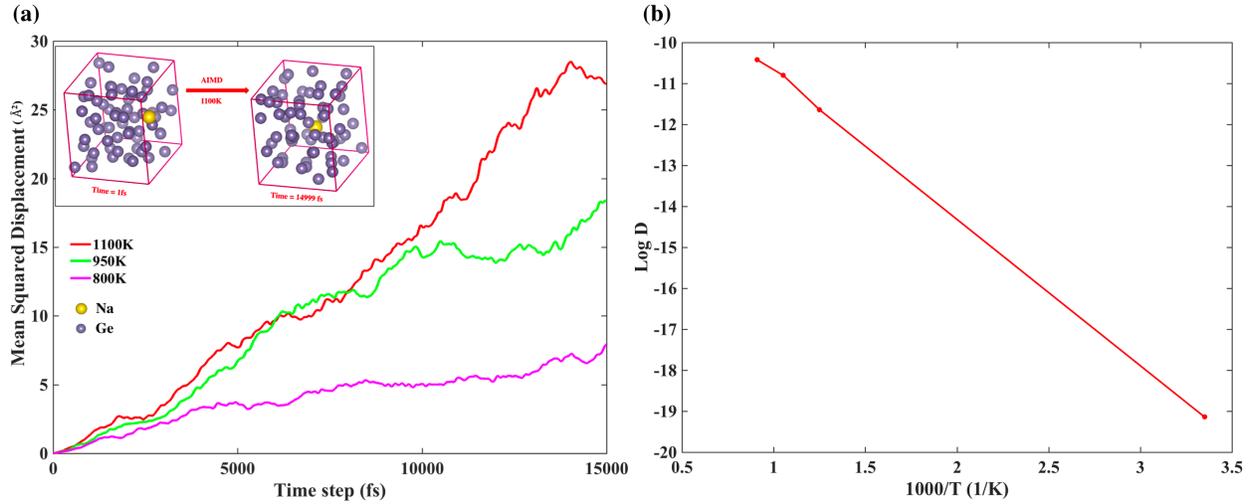

*Figure 5*: (a) MSD plot for Na in *a*-Ge with respect to time corresponding to different temperatures namely 1100K (red), 950K (green) and 800K (magenta). (b) Arrhenius plot of log of diffusivity vs inverse of temperature (1/K) for diffusion of Na in *a*-Ge extrapolated to room temperature.

*Table 2*: Mean Squared Displacement and calculated diffusivities for various temperatures of Na in Ge.

| Temperature (K) | MSD (Å$^2$) | Diffusivity (cm$^2$/s) |
| --- | --- | --- |
| 1100 | 26.89 | 29.87 X 10$^{-6}$ |
| 950 | 18.3413 | 20.37 X 10$^{-6}$ |
| 800 | 7.928 | 8.80 X 10$^{-6}$ |

## 4. CONCLUSION

In conclusion, we investigated sodiation kinetics in *a*-Ge anode which is least explored alloying element of group 14 but has potential to be a promising anode for NIB due to its similarities to Li-Si system. Moreover, Ge electrode yields high theoretical capacity of 369 mAhg$^{-1}$. We analyzed the intercalation potential and capacity correlation for intermediate equilibrium structures and our computational results are in excellent agreement with the existing experimental data. Our equilibrium curve lies in middle potential values of experimental curve. From our first principle results, we also computed the volume expansion of Na-Ge alloy electrode to be approximately 149.51% in the fully sodiated state ($Na_{64}Ge_{64}$). It is well established that in Na-Ge battery system, starting with *a*-Ge, upon complete sodiation results in an amorphous system ($Na_{64}Ge_{64}$). However, not much information about the intermediate stages of Na-Ge system exists. In the present study, intermediate Na concentrations (6%, 19%, 31%, 50%, 69% and 88%) in Na-Ge system were assessed to identify any possible phase change during sodiation. We found that in spite of starting and final stable amorphous phases, system undergoes minute phase transitions to crystallinity for smaller Na concentrations (Na < 20%). This information was revealed in PCF analysis of Na-Na pairs in the system. While PCF of Ge-Ge pairs showed amorphous nature throughout, we observed peaks referring to crystallinity in Na-Na PCF plot for Na concentration below 20%. It was noted that after 50% sodiation, system was amorphous throughout. Moreover, we studied diffusivity of Na in *a*-Ge and found the energy barrier for diffusion of Na in *a*-Ge to be much lower as compared to crystalline Ge (*c*-Ge). We calculated diffusivity of single Na in *a*-$Ge_{64}$ to be $4.876 \times 10^{-9}$ $cm^2/s$ at 300 K, which is greater than previously reported diffusivities of Na in other group 14 and 15 elements. Our systematic investigation yields in-depth insight into the sodiation kinetics and provides guidelines for experimentalists for optimal design of Ge-based NIB for real-life applications.

## Acknowledgement

DD acknowledges NJIT for the faculty start-up package. We thank Prof. Siva Nadimpalli of NJIT for his suggestion throughout the project. We are grateful to the High-Performance Computing (HPC) facilities managed by Academic and Research Computing Systems (ARCS) in the Department of Information Services and Technology (IST) of the New Jersey Institute of Technology (NJIT). Some computations were performed on Kong.njit.edu HPC cluster, managed by ARCS. We acknowledge the support of the Extreme Science and Engineering Discovery Environment (XSEDE) for providing us their computational facilities (Start Up Allocation - DMR170065 & Research Allocation - DMR180013). Most of these calculations were performed in XSEDE SDSC COMET Cluster.